\documentclass[12pt]{iopart}
\pdfoutput=1
\usepackage[final]{graphics}
\usepackage{iopams}
\usepackage{amssymb}
\usepackage{amsfonts}
\usepackage{epsfig}
\usepackage{graphicx}
\usepackage{color}
\usepackage{subfigure}

\begin{document}
\title[Leal et al., Kinetic modelling of epitaxial film growth]
{Kinetic modelling of epitaxial film growth with up- and downward step barriers}
\author{F.F. Leal$^{1,2}$, T. J. Oliveira$^1$ and S. C. Ferreira$^1$\footnote{On leave at Departament de 
F\'{\i}sica i Enginyeria Nuclear, Universitat Polit\'ecnica de Catalunya,
Spain.} 
}

\address{$^1$Departamento de F\'{\i}sica - 
Universidade Federal de Vi\c{c}osa, 36571-000, Vi\c{c}osa,
Minas Gerais, Brazil}
\address{$^2$Instituto Federal de Ci\^encia, Educa\c{c}\~ao e Tecnologia - 
Campus Itaperuna, Rodovia BR 356 - KM 03, 28300-000, Itaperuna, 
Rio de Janeiro, Brazil.}

\ead{ffleal@iff.edu.br,silviojr@ufv.br,tiago@ufv.br}

\begin{abstract}
The formation of three-dimensional structures during the epitaxial growth of
films is associated to the reflection of diffusing particles in descending
terraces due to the presence of the so-called Ehrlich-Schwoebel (ES) barrier. We
generalize this concept in a solid-on-solid growth model, in which a barrier
dependent on the particle coordination (number of lateral bonds) exists whenever
the particle performs an interlayer diffusion. The rules do not distinguish
explicitly if the particle is executing a descending or an ascending interlayer
diffusion. We show that the usual model, with a step barrier in descending
steps, produces spurious, columnar, and highly unstable morphologies if the
growth temperature is varied in a usual range of mound formation experiments.
Our model generates well-behaved mounded morphologies for the same ES barriers
that produce anomalous morphologies in the standard model. Moreover, mounds are
also obtained when the step barrier has an equal value for all particles
independently if they are free or bonded. Kinetic roughening is observed at long
times, when the surface roughness $w$ and the characteristic length $\xi$ scale
as $w\sim t^\beta$ and $\xi\sim t^\zeta$ where $\beta\approx 0.31$ and
$\zeta\approx 0.22$, independently of the growth temperature.
\end{abstract}

\pacs{05.70.Ln, 89.75.Da, 89.75.Hc, 05.70.Jk}

\submitto{\it J. Stat. Mech.: Theor. Exp.}

%\maketitle

\section{Introduction}

Since epitaxial techniques became available, a large number of experimental
works reporting on the production of films with three-dimensional patterned
structures, called generically of mounds, as well as the analytical and
computational modelling devoted to explain these patterns have been found in
literature (see Ref.~\cite{Evans2006} for a review).  Mounds have been observed
during the growth of a wide diversity of films, ranging from
metals~\cite{Jorritsma,Caspersen,Yon2010} to inorganic
\cite{Johnson,EslaniPRL2006} and organic~\cite{ZorbaPRB2006,Hlawacek2008}
semiconductor materials. The morphological properties, one of the most important
features in film production, depend of distinct physical mechanisms as diffusion
rate~\cite{krugbook,Evans2006}, deposition rate~\cite{Hamouda2008}, growth
temperature~\cite{Ferreira2006.APL,EslaniPRL2006}, deposition
shadowing~\cite{AmarPRB2008}, among others.

During the film production, the surface may follow distinct growth
regimes~\cite{Evans2006}. The so-called kinetic roughening, where the surface
dynamics is described by scale invariance, has significant interest. The power
law behaviours of the surface rms width $w\sim t^\beta$ and the characteristic
correlation length $\xi\sim t^{\zeta}$ are the basic scaling laws involved in
the kinetic roughening, defining the growth and coarsening exponents,
respectively. There is a third, not independent exponent related to the scaling
of the global surface fluctuations, the roughness exponent $\alpha =
\beta/\zeta$~\cite{Ramasco}. Two scaling regimes are of special interest in
mound formation: The first one involves slope
selection~\cite{vvdenskyPRB1995,SiegertPRL1994,SchinzerEJB2000}, where the ratio
between the characteristic length and the surface width approaches a constant
value. This condition  is satisfied for $\beta=\zeta$. The second one is the
super-roughness scaling regime~\cite{Lopez1997,Ramasco}, characterized by
$\alpha>1$. The latter implies that the surface is locally smooth, having
different scaling exponents for global and local fluctuations~\cite{Lopez1997}.

The mound formation during the epitaxial growth is frequently associated to the
reflection of particles (atoms or molecules) in descending
terraces due to presence of the so-called Ehrlich-Schwoebel (ES)
barrier~\cite{Evans2006}. The explanation for this step barrier is founded in the
experimental observation that particles are reflected backwards in the terrace edges
more frequently than move to a lower
layer~\cite{schwoebel:1966,ehrlich:1966}. Although other mechanisms, as
short-range attraction towards ascending steps~\cite{Amar1996} and fast edge
diffusion~\cite{Murty2003,Chatraphorn,Pierre-Louis} can lead to mound formation,
they are weaker than the ES barrier.

Kinetic Monte Carlo (KMC) is a standard simulation method for atomistic
modelling of epitaxial growth. The basic idea was formerly introduced by  Clarke
and Vvedensky \cite{Clarke,ClarkeJAP} and generalized in many other
frameworks~\cite{Evans2006,krugbook}. In the Clarke-Vvedensky model, diffusion
is a thermally activated process described by an Arrhenius law $D\sim
\exp(-E/k_BT)$~\cite{krugbook}, where $E$ is a diffusion activation energy,
$T$ the temperature and $k_B$ the Boltzmann constant. The ES barrier is usually
modelled with an additional energy barrier $E_b$ to particles moving downwards
in the edge of a terrace~\cite{Evans2006}. In a simplified picture, $E_b$ is the
additional energy needed when a particle crosses a step edge. Therefore, a step
barrier must also exist for particles performing upward diffusion (see
Fig.~\ref{fig:model1}). However, the ES barrier is generally considered only
for particles diffusing from an upper to a lower terrace and barriers for
ascending steps were considered in a limited number of models. \v{S}milauer and
Vvedensky~\cite{vvdenskyPRB1995} proposed a step barrier proportional to the
difference between next-nearest neighbour (NNN) bonds before and after the
particle hop only if the number of NNN bonds decreases after the hop,
independently if the adatom performs an up- or downward diffusion. Recently, a model with a
barrier dependent on the step height was investigated~\cite{Leal2011}. Besides
the barrier in descending steps, this model also assumes an ascending barrier
for multilayer steps. However, the barrier in monolayers was implemented as
usual and the model is not fully symmetric in to relation down- and upward step
diffusion. Barriers in ascending steps were also considered in the limited mobility
model of Wolf-Villain~\cite{Rangdee2006}, where probabilities of moving to lower
or upper terraces after a deposition step were introduced. Mound formation is
obtained when the probability of upward diffusion is larger than downward
diffusion.

In the present work, we revisit the effects of ES barriers in KMC simulations by
comparing the standard model, having a barrier only in descending steps, with
the case where a barrier for upward diffusion is also activated. Moreover, we
consider different barriers depending if the particle has or not lateral bonds.
The model allows to investigated the interesting symmetric case where bonded and
free (with no lateral bonds) particles undergo the same barrier. We show that
the standard bond counting model produces unrealistic columnar mounds when the
diffusion of particles with lateral bonds cannot be neglected. This feature
limits the applicability of this model to a short range of temperature. On the
other hand, the model with a bond dependent step barrier produces well-behaved
mounded surfaces in a wide range of temperature, consistent with typical
experimental conditions.

We have organized the paper as follows. In Sec.~\ref{sec:model} we present the
models and KMC simulation strategies. Comparison between a barrier exclusively
to downward diffusion and the bond dependent step barrier is done in
Sec.~\ref{sec:result}. Further analyses of our model, including the symmetric
case, are presented and discussed in Sec.~\ref{sec:BDSB}. We summarize our
results and conclusions in Sec.~\ref{sec:conclu}. 

\section{Kinetic Monte Carlo model}
\label{sec:model}
We model homoepitaxy on a substrate at a temperature $T$ represented by a
triangular lattice with periodic boundary conditions. Particles are deposited at
a constant rate $F$ under the solid-on-solid restriction, implying that voids
and overhangs are absent. A deposition event consists in increasing the height
by $h_j\rightarrow h_j+1$, where $h_j$ is the number of particles deposited in
the site $j$. The choice of the deposition site $j$ is done in two steps:
Firstly a site is chosen at random, and secondly a new particle is deposited in
the most energetically favourable (largest bond number) site among the chosen
one and its nearest neighbours (NN). This transient mobility is justifiable if
we consider that the particle arrives at the substrate with a kinetic energy
higher than the typical activation energy in the substrate ($k_{B} T$). This
deposition rule corresponds to the classical Wolf-Villain (WV) model~\cite{WV}.
The transient mobility in the deposition rule plays an import role
in the surface morphology at low temperatures, as we will show in Sec.
\ref{sec:BDSB}. It is important to mention that a transient mobility was used in
previous models of epitaxial growth with thermally activated
diffusion~\cite{Evans2006,vvdenskyPRB1995}.

Intra and interlayer thermally activated diffusion are allowed for any surface
particle. Diffusion happens between NN sites and depends on the initial and
final position of the particle. The diffusion rate from site a $j$ to a neighbour
$j'$ is given by the Arrhenius law 
\begin{equation}
\label{eq:difrate}
 D(j,j';T) = \nu\exp\left[-\frac{E(j,j')}{k_B T}\right],
\end{equation}
where $\nu$ is an attempt frequency and $E$ the diffusion activation energy
given by
\begin{equation}
\label{eq:energ}
 E(j,j') = E_0+n_jE_N+E_b(j,j').
\end{equation}
The energy $E_0$ represents the interaction with the substrate and/or bulk of
the film, $E_N$ is the contribution of each one of $n_j$ lateral bonds, and
$E_b$ is the step barrier present for interlayer diffusion. Apart of the
step-edge barrier, the model is equivalent to the thermal activation model
proposed by \v{S}milauer  and Vvedenski~\cite{vvdenskyPRB1995}. Algorithmically,
the step barrier is present only if $\Delta h(j,j') \ne 0$, where $\Delta
h(j,j') \equiv h_{j'}+1-h_j$.

Apart from the details of the barrier rules, which  are described below, the
kinetic Monte Carlo simulations are implemented as follows. A particle is
deposited and then $N$ diffusion attempts are sequentially implemented.
In a diffusion attempt, a particle of the surface is selected
at random and its diffusion is implemented accordingly to the ratio
$D/F$ given by Eq.~(\ref{eq:difrate}). The diffusion is efficiently implemented
by keeping updated lists containing the positions of the particles with the same
number of lateral bonds. Let $\rho_n$, $n=0,1,\ldots,6$ be the fraction of
particles with $n$ lateral bonds. Between two deposition events, the average
number of diffusion attempts of the particles having $n$ bonds is
$N_n=\rho_nD(n;T)/F$ where 
\begin{equation}
D(n;T)=\nu\exp\left(-\frac{E_0+nE_N}{k_BT}\right)
\label{eq:dn}
\end{equation}
is the diffusion rate excluding the step barrier contribution. In the
implementation, we sequentially choose $N=\sum_{n=0}^6 N_n$\footnote[1]{More
precisely, the number of attempts is increased by 1 with probability
$N-\bar{N}$, were $\bar{N}$ is truncation to the integer part of $N$} particles,
with the coordination number obeying the ratio $N_0:N_1:\cdots:N_6$. For each
selected particle (site~$j$), a neighbour (site~$j'$) is chosen with equal
chance and the particle hops to this site with probability 
\begin{equation}
\label{eq:Ph}
 P_{h} (j,j';T) = \exp\left[-\frac{E_b(j,j')}{k_BT}\right],
\end{equation}
which includes the effect of the step barrier. With the complementary
probability the  particle stays in the same site. The process is repeated until
the runs of $N$ diffusion attempts are finished and a new particle deposited. 

\begin{figure}[t]
 \centering
 \includegraphics[width=9cm]{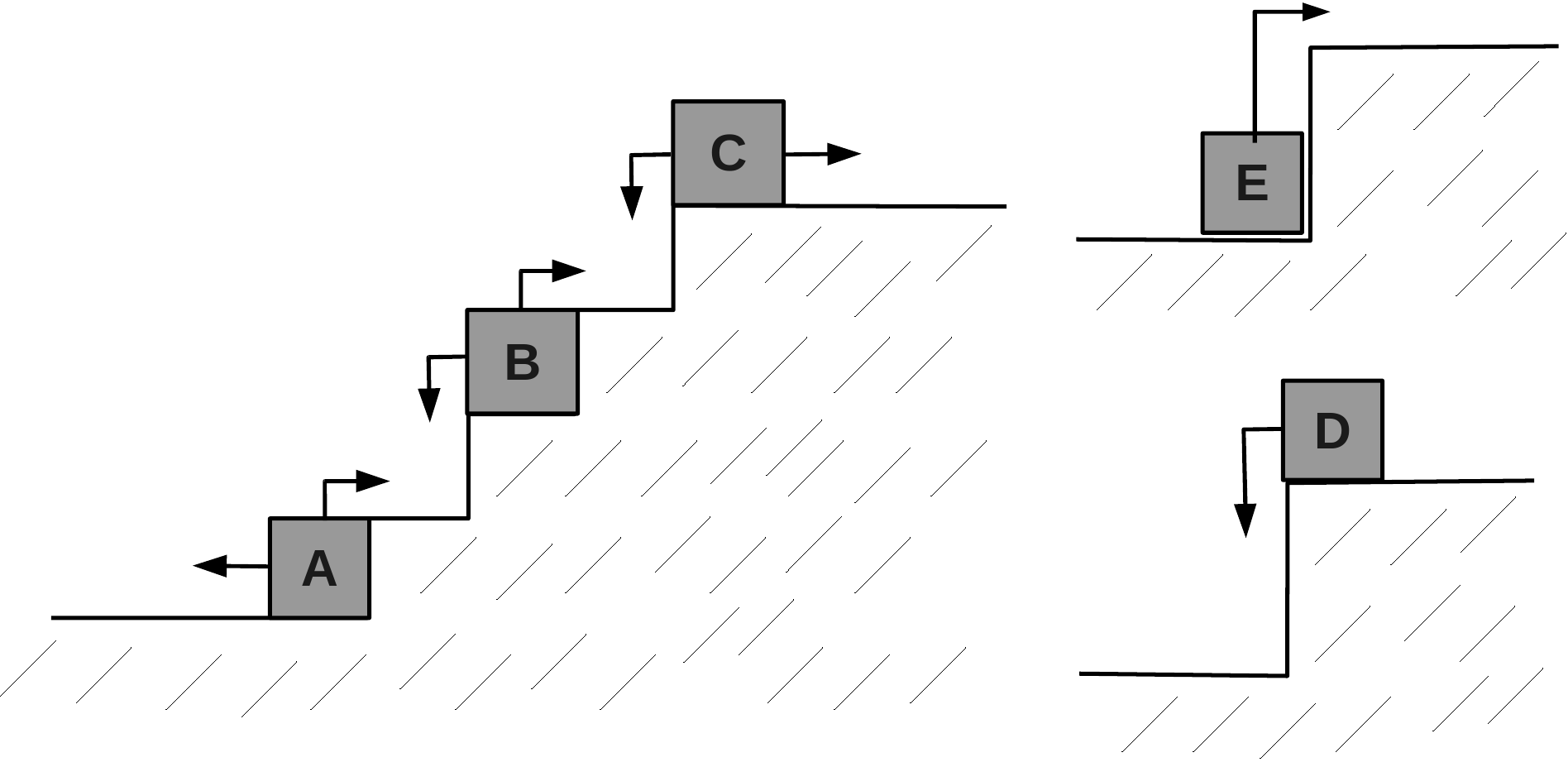}
 % diffusions.pdf: 351x215 pixel, 72dpi, 12.38x7.58 cm, bb=0 0 351 215
 \caption{Illustration of important situations involving interlayer diffusion:
Diffusion through monolayer (left) and multilayer (right) steps. Both, particles
\textit{D} and \textit{E} have to overcome a barrier when moving through the 
steps.}
 \label{fig:model1}
\end{figure}

In the present work, we are interested in the effect of an uphill in addition to
the usual downhill step-edge barrier. Figure~\ref{fig:model1} shows some important
situations involving interlayer diffusion. Let us firstly consider the particles
\textit{A} and \textit{C} lying in monosteps. In both cases, it is intuitive
that staying in the same layer is favoured since, otherwise, the particle has to
detach and rebind in an adjacent layer. So, an additional step barrier (not
necessarily equal) must be present in both configurations. The kink particle
\textit{B}, however, should diffuse up- or downwards with equal chance due to
the local symmetry of this configuration. In a simple description, particles
\textit{A} and \textit{B} may be subject to a similar step barrier that may
possibly be different from the barrier for particle \textit{C}. 

%In the present work, we are interested in the effect of an uphill in addition to
%usual downhill step-edge barrier. In fact, considering that the origin of the
%barrier is the reflection of the diffusing adatom near a step-edge, the particle
%$E$ (in Fig.~\ref{fig:model1}) must be subject to the same effects that the
%particle $D$, therefore, we may expect an ES-like energy barrier also for that
%particle. In the same way, particles lying in a kink (particle $B$) must have
%the same probability to move up- or downwardly, given the local surface
%symmetry, i. e., again the energy barrier should be the same. However, in the
%last case, the reflection of the particle for the step-edge does not make sense,
%since the particle is part of the proper edge. Thus, it is reasonable consider
%different energy barriers for kink particles ($A$ and $B$) and free adatoms ($C$
%and $D$).

Our model has descending and ascending interlayer diffusions that depend if the
particles are  bonded or not. By simplicity, we assume two kind of step
barriers: $E_{b0}$ for free particles ($n=0$) and $E_{bn}$ for bonded particles
($n\ge1$). Notice that rules do not concern explicitly if the particles are
moving down- or upwardly, but, free particles will never be in an ascending step
since this configuration requires $n\ge1$. In principle, configurations
\textit{D} and \textit{E} are apparently equivalent but the situation is much
more complicated in 2+1 dimensions due to the bonds not shown in an 1+1
representation. However, it is clear that a step barrier must exist for both
particles and our model is, therefore, an improvement for these configurations.
Actually, as shown in Ref. \cite{Leal2011}, although the presence of a barrier
in a multilayer step is important, the details of the barrier dependence with
the step height do not alter significantly the final surface morphology. We
refer to our rules as BDSB (bond dependent step barrier) while the standard
model we call of DSB (downward step barrier).

\section{Results: BDSB versus DSB models}
\label{sec:result}

\begin{figure}[hbt]
 \centering
 \includegraphics[width=12cm]{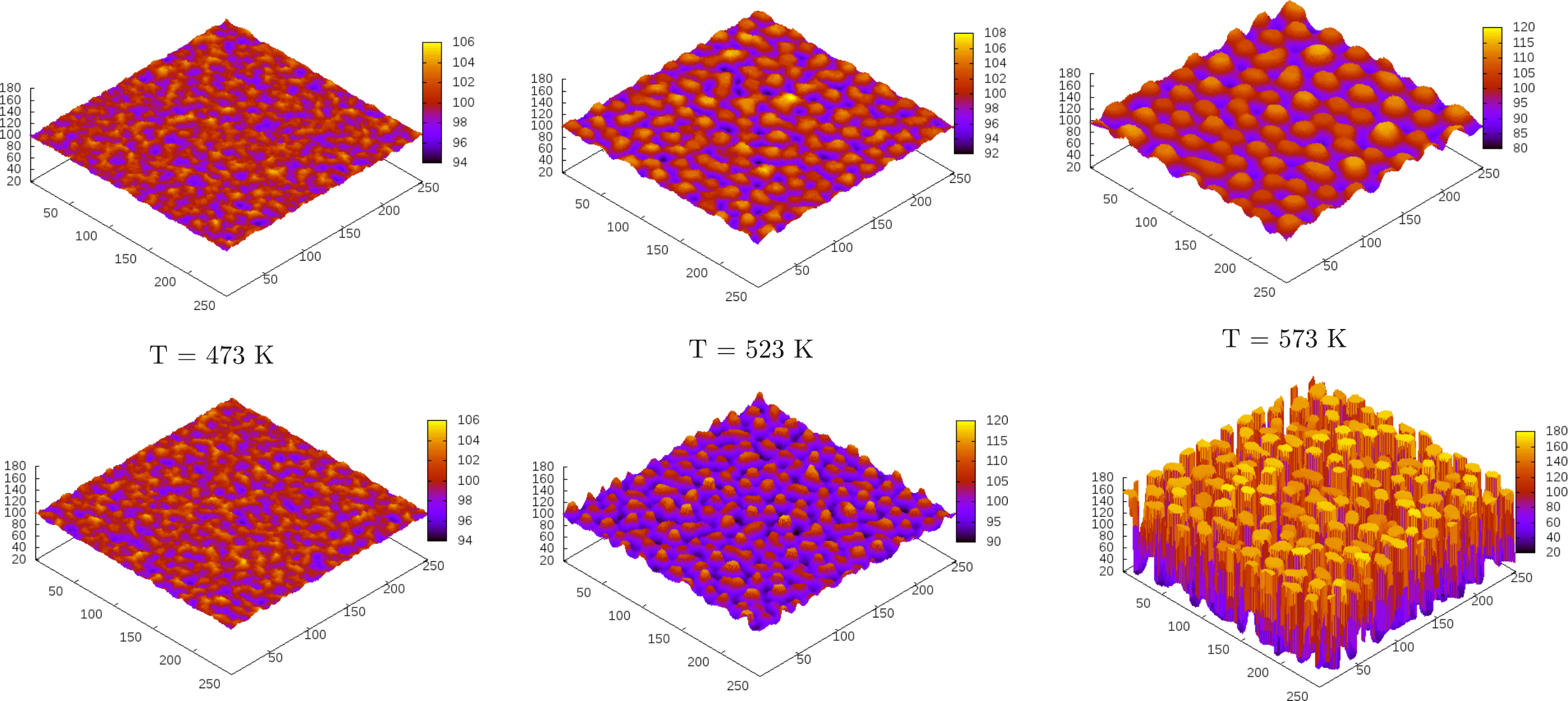}
 % diffusions.pdf: 351x215 pixel, 72dpi, 12.38x7.58 cm, bb=0 0 351 215
 \caption{Surface morphologies for the BDSB (top) and the DSB (bottom) 
 models for distinct temperatures after a deposition of 100~ML. 
 The step barriers in the BDSB model are $E_{bn}=0$ and 
 $E_{b0}=0.05$~eV while the barrier in the DSB 
 model is $E_b=0.05$~eV. See supplementary material for surface evolutions.}
 \label{fig2}
\end{figure}

The KMC simulations were performed with the fixed parameters $E_0=1$~eV,
$E_{N}=0.11$~eV, $F=1$ monolayers (ML) per second, and $\nu=10^{13}$~s$^{-1}$.
Temperature was varied in the interval  $T=473-623$~K corresponding to a range of
temperature $\Delta T=150$~K compatible with several semiconductor and metal film
growths~\cite{Evans2006}. We simulated substrates sizes $L=256$, $512$ and $1024$, and no
significant difference was observed in the results. Therefore, except if explicitly mentioned,
the results correspond to $L=512$. 

In order to analyse the effects of the step barrier for kink particles, we
compare the DSB and BDSB models using a null barrier for bonded particles
($E_{bn}=0$). In this case, BDSB includes an ES barrier only for free particles
in descending steps  while in DSB model the ES barrier is always present
in downward movements. Figure~\ref{fig2} shows the surface morphologies for three
temperatures after a deposition of 100~monolayers. A fixed step barrier of
0.05~eV is used for free particles in BDSB model and for downward diffusion in
DSB model. Both models exhibit mounded morphologies, qualitatively similar at
low temperatures (see $T=473$~K, for example). However, when temperature is
increased, the surfaces of DSB model become anomalously mounded, with
unrealistic columnar structures and  huge steps while regular mounded
morphologies are observed for BDSB. If the deposition time is increased, the
surface for $T=523$~K also develops columnar mounds in DSB model, but does not
for BDSB, as can be seen in the videos 1 and 2 of the supplementary material.
Increasing step barrier, representing more realistic values in many systems,
strongly enhances the anomaly in the DSB whereas BDSB remains well-behaved (data
not shown).

The DSB unrealistic behaviour at high temperatures has a simple explanation
formerly realized by Villain~\cite{Villain}: For a particle in a descending step
(particle C, Fig.~\ref{fig:model1}), the chances of attachment in an ascending
step or of a new terrace nucleation  in the same layer are greater than the chances
of attachment in the subjacent step due to ES barrier. However, at high
temperatures, when a particle in an ascending step (particle A,
Fig.~\ref{fig:model1}) has a non-negligible mobility, the detachment and
climbing chances are equal and the asymmetry of the rule for kink
particles (particle B, Fig.~\ref{fig:model1}) produces a large destabilizing
uphill current. This result shows that the step barrier in the kink particles
plays a central role in the determination of the mound morphology.

%In the case of a ES barrier only in
%the descending steps, a large barrier for kink particles ($E_{bn} \approx
%E_{b0}$) leads to columnar structures in surfaces, while a small one ($E_{bn}
%\ll E_{b0}$) produces regular patterns consistent with experiment outcomes.
%and suggests that
%different step barriers for kink and free adatoms may exist in real systems. 

\begin{figure}[t]
 \centering
\subfigure[\label{fig3a}]
{\includegraphics[width=7cm]{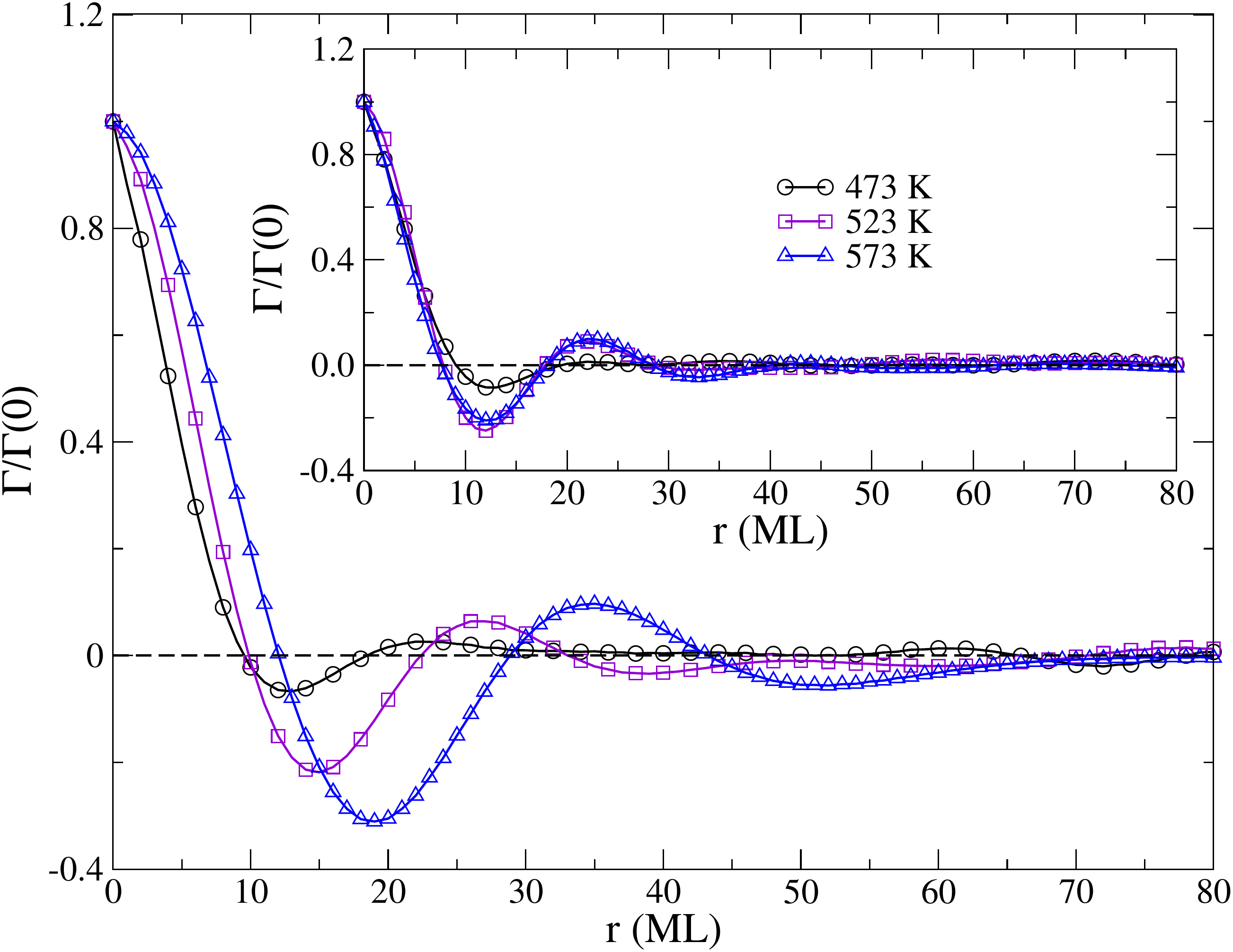}} 
\subfigure[\label{fig3b}]
{\includegraphics[width=7cm]{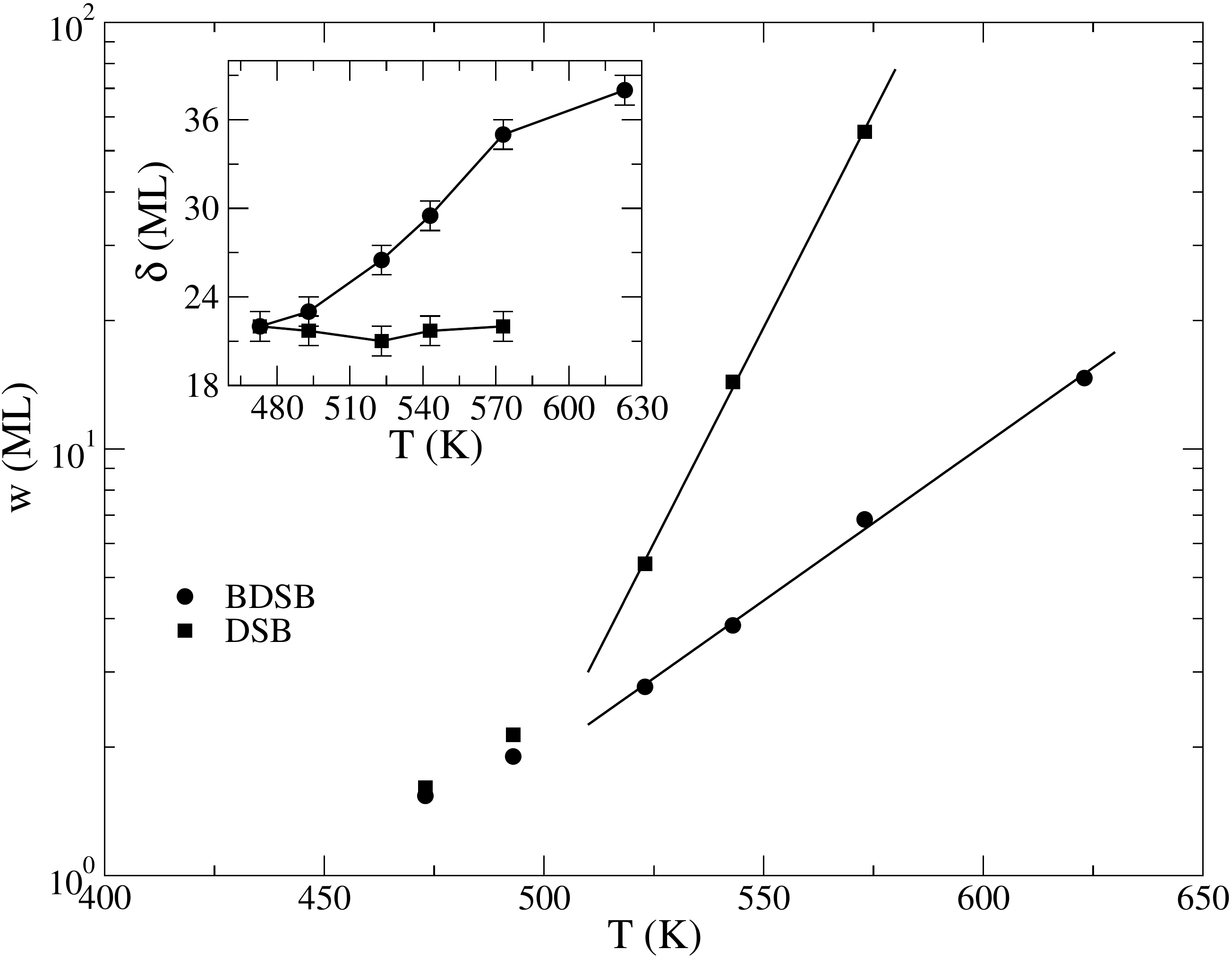}}
\caption{Surface morphological characterization for BDSB and DSB models. (a)
Height-height correlation function after the deposition of 100~ML at distinct
temperatures. Results for BDSB and DSB models are shown in the main plot and 
inset, respectively. (b) Interface width (main plot) and characteristic mound separation
(inset) against temperature. The fixed parameters are the same used in
Fig.~\ref{fig2}.}
 \label{fig3}
\end{figure}

A basic quantity used to characterize the surface morphology is the 
height-height correlation function defined as~\cite{Murty2003,Chatraphorn}
\begin{equation}
\label{eq:cor}
\Gamma(\mathbf{r}) = 
\langle h(\mathbf{x})h(\mathbf{x}+\mathbf{r})\rangle_\mathbf{x}, 
\end{equation}
where $h(\mathbf{x})$ is the surface in the mean height reference and
$\langle\cdots\rangle_\mathbf{x}$ is the average over the surface. The interface
width (rms roughness) $w$ is given by $\sqrt{\Gamma(0)}$ and the first maximum
of $\Gamma(r)$ measures an average distance between mounds $\delta$.
Figure~\ref{fig3a} shows the height-height correlation function against distance
for systems at different temperatures. The curves were rescaled by $\Gamma(0)$ to improve
visibility. The oscillating correlation function is the hallmark of mounded
surfaces~\cite{Evans2006,Murty2003}. Figure~\ref{fig3b} shows the surface width
and mound separation against temperature for both models. The interface width
grows exponentially with temperature in both cases meaning a temperature driven
instability, but the DSB produces anomalously huge surface widths in the
investigated temperature interval. The characteristic mound distance is closely
constant for DSB and grows monotonically with
temperature for BDSB, as shown in the inset of Fig.~\ref{fig3b}.

It is important noticing that DSB model produces a closely constant mound separation
and a fast increase of the surface width with temperature. Morphologically, we
have a strong increase of the mound heights and grooves. At higher temperatures
(not shown), the uphill currents are so intense that parts of the initial
substrate remain uncovered, and highly columnar structures are completely formed
upon the template of sub-monolayer islands.

\begin{figure}[t]
 \centering
 \includegraphics[width=13cm]{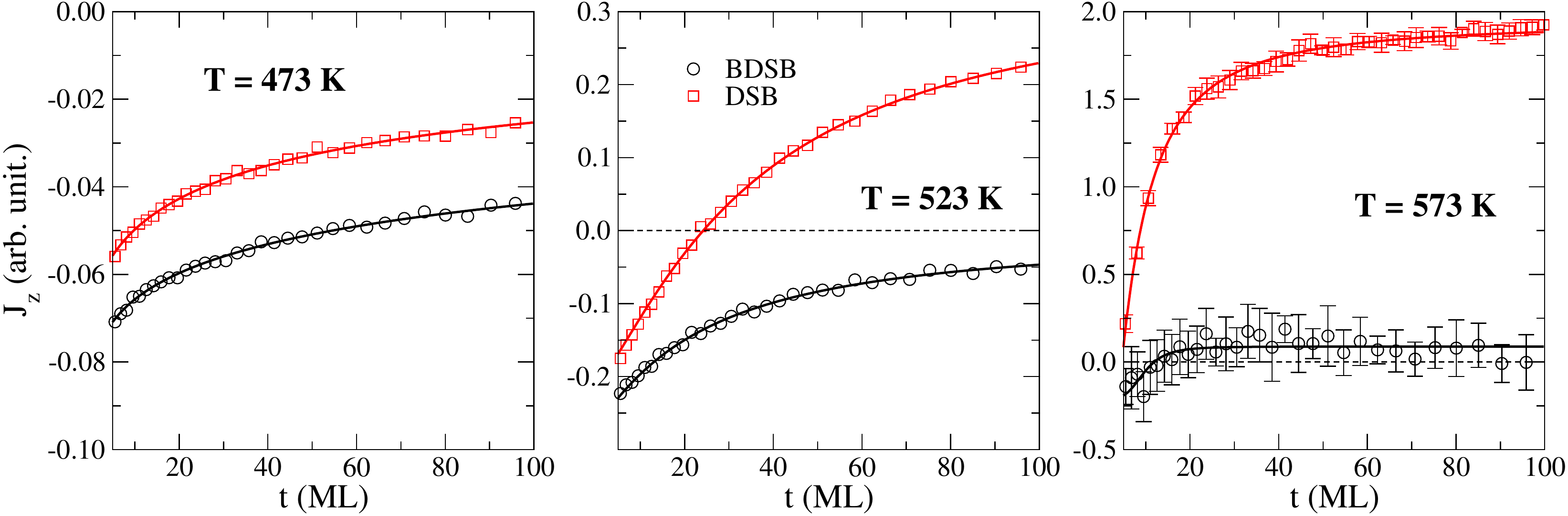}
 % standvsnosso_subs.pdf: 462x314 pixel, 72dpi, 16.30x11.08 cm, bb=0 0 462 314
\caption{Out-of-plane currents against time for BDSB and DSB models at distinct
temperatures. Step barriers are the same used in simulations shown in
Fig.~\ref{fig2}. Symbols represent simulations and solid line sigmoid
regressions as guides to the eyes. The data correspond to averages over 50
samples  and the error bars smaller than symbols were omitted.}
 \label{fig:cur1}
\end{figure}

The growth of three-dimensional structures involves a complex relation between the
down- and upward fluxes through the steps in addition to the adsorption and
nucleation mechanisms. We quantify the net flux though steps using an
out-of-plane current defined by
\begin{equation}
\label{eq:jz}
J_z = \frac{1}{2L^2}\sum_j D(n_j;T) \frac{1}{q} {\sum_{j'\in\mathcal{V}(j)} 
\Theta[\Delta(j,j')] P_{h}(j,j';T)},
\end{equation} 
in which the sum over $j$ runs over all sites  while the sum over $j'$ runs over all
$q$ nearest neighbours of $j$. The factor $1/2$ accounts for the double counting
of bonds $(j,j')$. The factor $D(n_j;T)$ is given by
Eq.~(\ref{eq:dn}), $P_{h}(j,j';T)$ by Eq.~(\ref{eq:Ph}), and $\Theta(x)$ returns
the sign of $x$ and $\Theta(0)=0$. This quantity is the average interlayer
diffusion rate per site since the intralayer diffusion is not counted. 
In order to reduce statistical fluctuations, the current at a time $t$ is the  average 
over a short interval $\delta t$ around $t$.
The currents for the BDSB and DSB models are compared in Fig.~\ref{fig:cur1}. For
low temperatures, the currents show a downward (negative) flux in both models
that is attenuated in the DSB due to the absence of barrier to upward diffusion.
The flux tends monotonically and slowly to zero. Apart of a short initial
transient, the data is very well fitted by a Hill function
\begin{equation}
J_z = -j_0+j_0\frac{t^\eta}{A+t^\eta}
\end{equation}
with exponents $\eta \approx 0.34$ and $\eta \approx 0.48$ for BDSB and DSB,
respectively. For intermediate temperatures, the BDSB current also grows
monotonically to a zero flux while a very large positive current is observed in
DSB. For high temperatures, both models have positive currents, but the currents 
in BDSB are much smaller than in DSB.

Slope selection is an important feature  in mound
formation~\cite{vvdenskyPRB1995,SiegertPRL1994,SchinzerEJB2000}, where the ratio
between characteristic width and height tends to a constant value at long times.
In terms of continuous equation approaches, the slope selection involves a
balance between up- and downhill currents~\cite{SiegertPRL1994,SchinzerEJB2000}
that is strongly violated in DSB model at intermediate and high temperatures, as
shown in Fig.~\ref{fig:cur1}. The current in the BDSB model at low and
intermediate temperatures, on other hand, grows monotonically to a balance
steady-state implying slope selection. The decaying and positive current
observed at high temperatures also states slope selection for
$t\rightarrow\infty$, but the time required to observe it is very large due to
a very slow decay (Fig. \ref{fig5b} shows a quantitative analysis for other
parameters). The absence of a characteristic slope in DSB model is also evident
in Fig.~\ref{fig2} due to the anomalous columnar morphology. Video 2 of the
Supplementary Material further illustrates the absence of slope selection in DSB
model.

\section{KMC simulations of the BDSB model}
\label{sec:BDSB}

\begin{figure}[t]
 \centering
\begin{minipage}{8.1cm}
\subfigure[\label{fig5a}]
{\includegraphics[width=8cm]{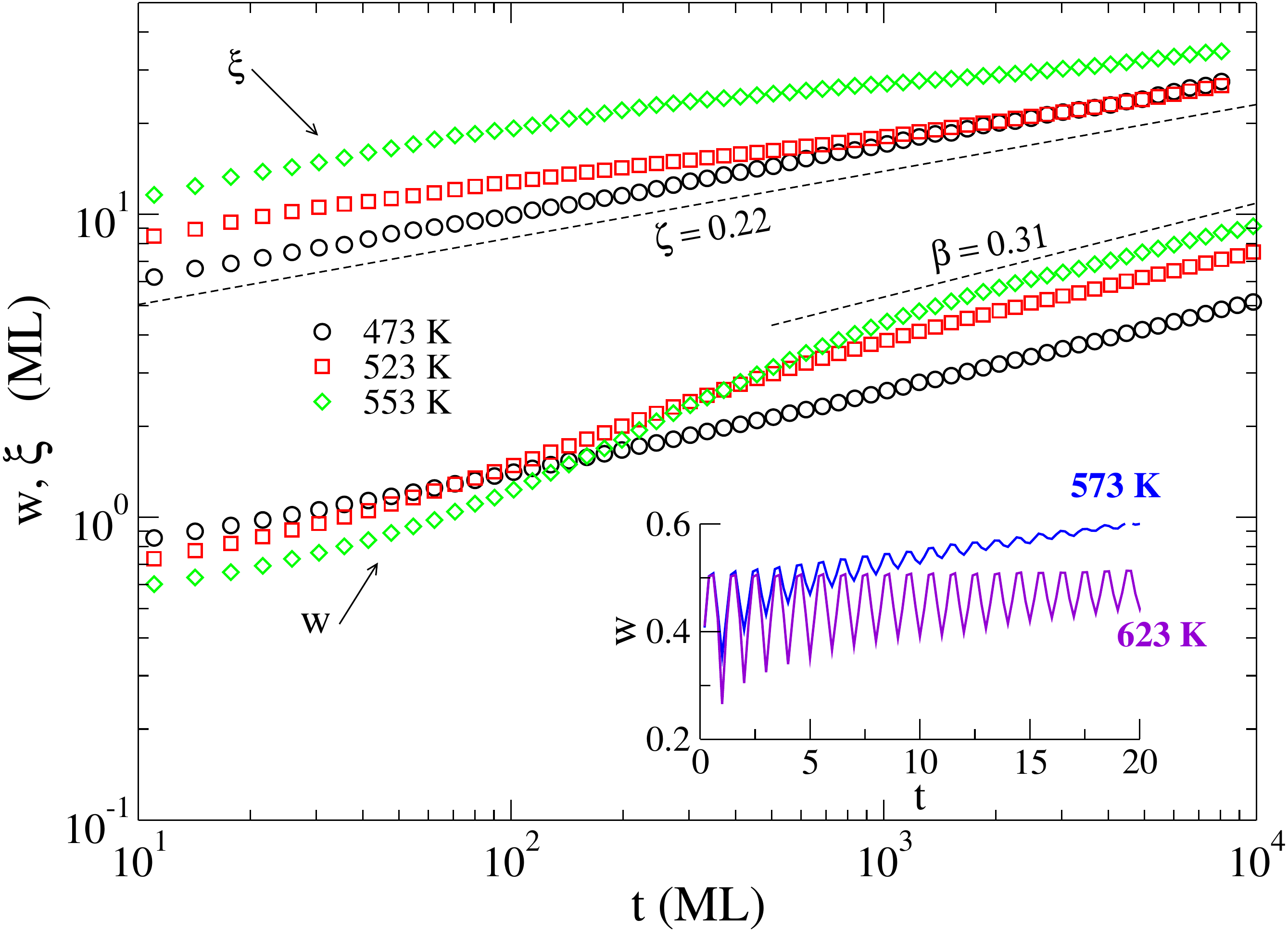}}  
\subfigure[\label{fig5b}]
{\includegraphics[width=8cm]{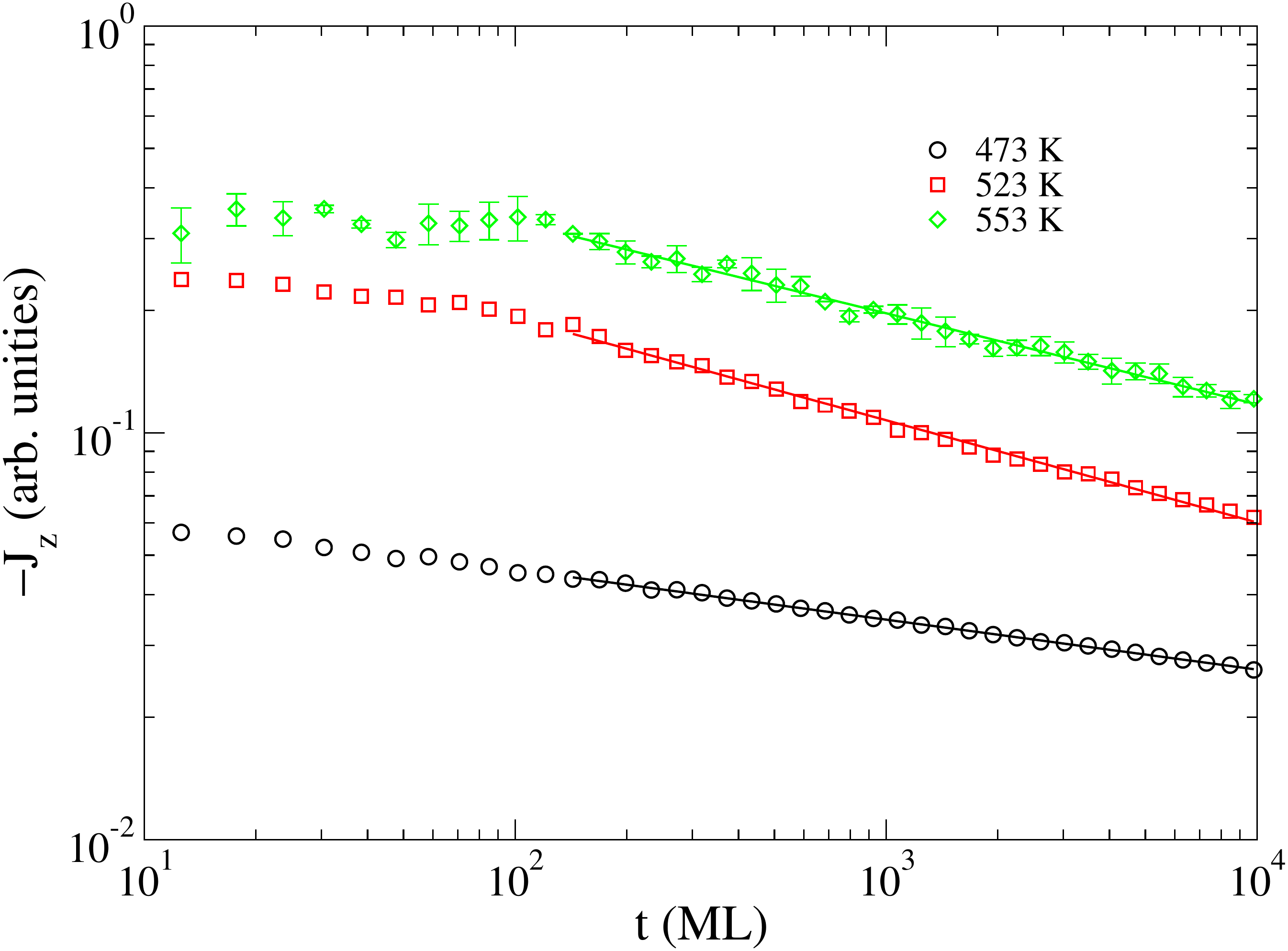}}  
\end{minipage}~
\begin{minipage}{5.1cm}
 \subfigure[\label{fig5c}]
 {\includegraphics[width=5cm]{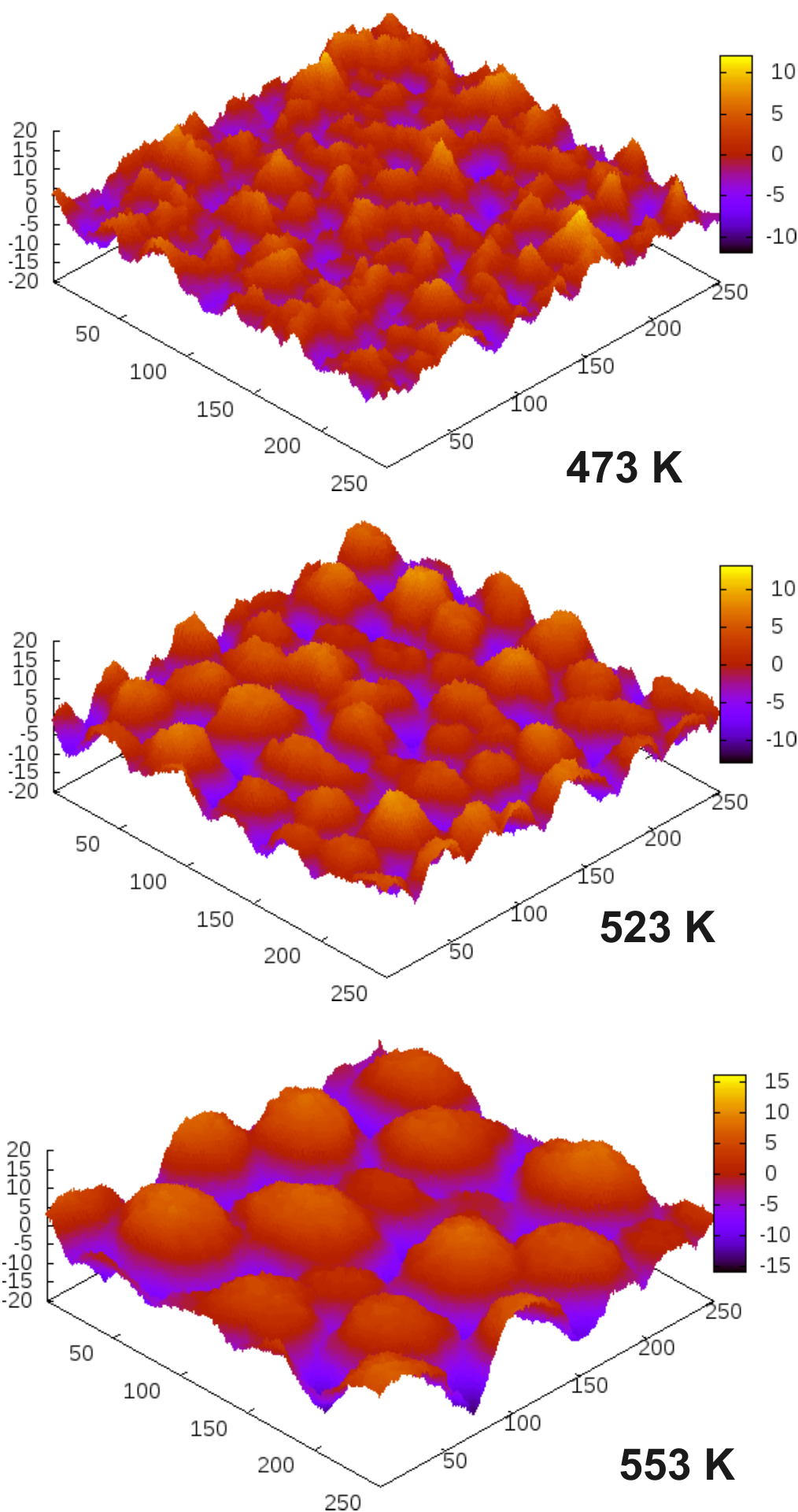}}
% surfEs0606_1000ML.pdf: 292x553 pixel, 72dpi, 10.30x19.51 cm, bb=0 0 292 553
\end{minipage}%\subfigure[\label{fig5c}]
%{\includegraphics[width=6.0cm]{./figs_mounds/PS0606T350.pdf}} 
 \caption{Surface dynamics for symmetric BDSB model with $E_{b0}=E_{bn}=0.06$~eV.
(a) Interface width and characteristic length against
time. Inset shows the interface width at short times. (b) Out-off plane
current against time. Solid lines are power regressions and error bars smaller
than symbols were omitted. (c) Surface morphologies after deposition of $10^3$ ML.}
 \label{fig5}
\end{figure}

Now, we concentrate in the effects of the uphill step barrier. The imbalance that
promotes the uphill destabilizing current and mound formation can be controlled with
the step barriers for free and bonded particles. Fig.~\ref{fig5a} shows the
surface width and characteristic length for the interesting symmetric case, in which
the barrier is the same for all particles, $E_{b0}=E_{bn}=0.06$~eV. The
interface width is strongly reduced if compared with the asymmetric barrier case
shown in Figs.~\ref{fig2} and \ref{fig3}. Therefore, an important role of the
uphill step barrier is the stabilization and smoothing of surfaces implying
reduction of the  three-dimensional structure amplitudes.

At high temperatures, the interface width presents an initial oscillating
behaviour of a layer-by-layer growth regime as shown in the inset of
Fig.~\ref{fig5a}. At longer times,  the layer-by-layer growth is replaced by a
kinetic roughening featured by asymptotic power laws $w\sim t^\beta$ and
$\xi\sim t^\zeta$, with exponents $\beta\approx 0.31$ and $\zeta = 0.22$. Here,
the characteristic length is defined as the first zero of height-height
correlation function~\cite{Murty2003,Chatraphorn}. The coarsening exponent
obtained in our simulations is close to that observed in a model with a barrier
dependent on the next nearest neighbours~\cite{vvdenskyPRB1995}. Our simulations
point out that the exponents are independent of temperature and barrier
strength, even though the asymptotic scaling regime of the characteristic length
was not reached in our simulations at high temperatures. The layer-by-layer
phase becomes longer for higher temperatures since, in these situations, the
particles have enough energy to overcome the edge barrier and to bind to highly
coordinated sites at the adjacent layers. The kinetic roughening was formerly
explained by Villain~\cite{Villain} in terms of the particle edge reflection
caused by the ES barrier to the downward movement, as discussed in
Sec.~\ref{sec:result}. For the symmetric barrier model, the origin of the
kinetic roughening is similar but the imbalance between downward and upward
currents is additionally because of particles in ascending steps (bonded by
definition) are, in an average, less diffusing than those in descending ones.
Notice that symmetric barriers hinder mound formation in the WV model in $d=1+1$
dimensions~\cite{Rangdee2006}. Mound formation was observed in the
WV model in $d=2+1$ without step-barrier, but the artefact of noise reduction was necessary to
overcome the very slow crossover to the asymptotic limit~\cite{Chatraphorn}.

Figure \ref{fig5a} also shows a re-entrant behaviour of the interface width
as function of temperature for intermediate growth times $t \simeq 100-300$~ML.
Re-entrant behaviours, which mean an interface width firstly increasing and then
decreasing with temperature (no monotonic dependence), were reported in a
classical experiment~\cite{Stoldt2000} and the corresponding kinetic
modelling~\cite{Bartelt1999,Yunsic2010} of the growth of Ag/Ag(100), among other
systems~\cite{Evans2006}. While specific models were devoted to describe this
complex behaviour~\cite{Evans2006,Bartelt1999,Yunsic2010}, it has appeared
spontaneously in our simulations during the crossover between layer-by-layer and
kinetic roughening growth regimes.

Interlayer currents against time for the symmetric barrier simulations are shown
in Fig.~\ref{fig5b}. The results  differ from the asymmetric barrier model since
the current is always negative and approaches zero as a power law
$J_z\sim-t^{-\eta}$ with exponents $\eta=0.1-0.25$. This algebraic decay means
that the balance between down- and uphill diffusion, theoretically required for
slope selection, will be reached only at infinite times when the
interface widths are already saturated. In fact, the growth and coarsening
exponents obtained in our simulations obey the relation $\beta>\zeta$ that
implies an aspect ratio of the mounds (height/width) increasing with time
and, consequently, no slope selection is strictly observed. A growth exponent greater
than the coarsening exponents also means that the surface is described by the so-called
super-roughness scaling regime~\cite{Ramasco} for which the roughness exponent
$\alpha = \beta/\zeta>1$. Generic scaling theory states that $\alpha>1$ implies
in locally smooth surface where the interface width in a scale of size
$\epsilon$ scales linearly as $w\sim \epsilon$~\cite{Ramasco}. 

Surface morphologies obtained with symmetric barriers at distinct temperatures
are shown in Figure~\ref{fig5c}, where one can clearly see three-dimensional
structures and mounds for the investigated temperature range. The mound shape is
approximately pyramidal for the lower temperature ($T = 473$~K) and becomes
dome-like for the higher ones. These finds are consistent with the theoretical claim
of a null current requirement for slope selection, since the pyramidal shape
with barely well-defined slopes was observed for the smallest current intensity
($T = 473$~K) while for the largest current intensity ($T = 553$~K), a signature of
slope selection is not evident.

\begin{figure}[bt]
\centering
 \subfigure[\label{fig6a}]
{\includegraphics[width=8.0cm]{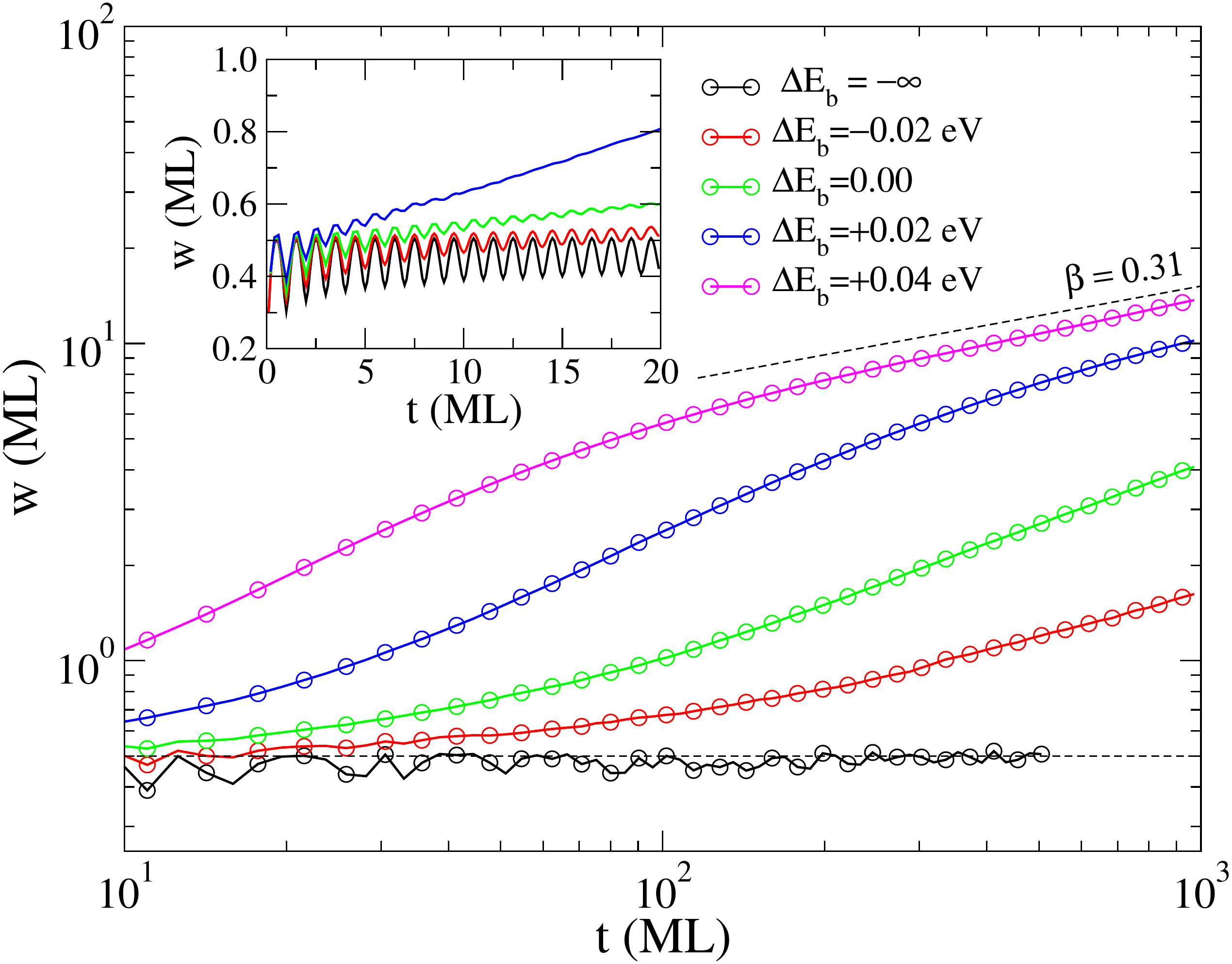}} 
\subfigure[\label{fig6b}]
{\includegraphics[width=12cm]{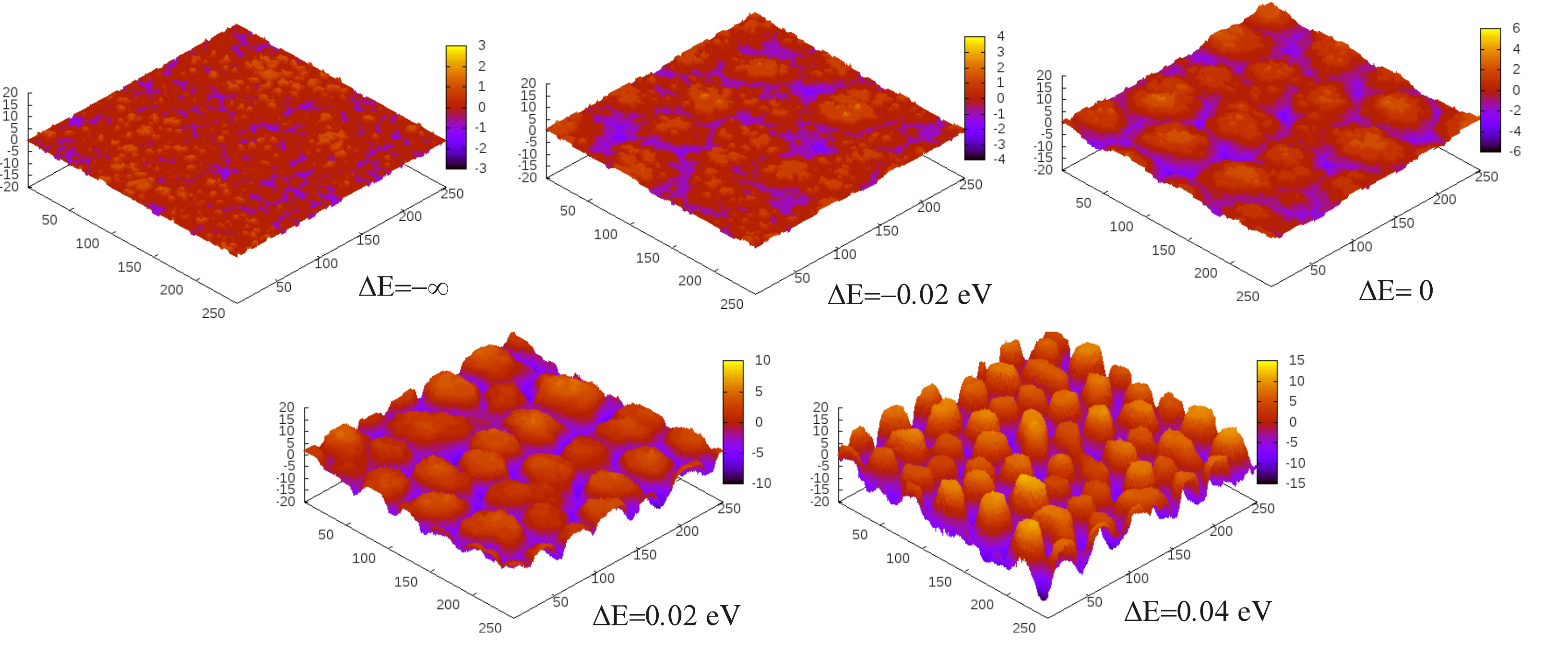}} 
\caption{(a) Interface width against time for fixed $T=573$~K and
$E_{b0}=0.06$~eV and distinct relative step barriers, $\Delta E_b =
E_{b0}-E_{bn}$, decreasing from top to bottom. The curves represent averages
over 10 independent samples. (b) Surface morphologies after the deposition of
100 ML.}
\end{figure}

The symmetric barrier is the simplest and, for this reason, the most
interesting case that produces mound formation. However, we can go beyond this
result and investigate the role of other mechanisms by exploring different
barriers for bonded and free particles, since there is no restriction
imposing equal barriers to both kinds of particles. The interlayer diffusion of
bonded particles also plays an essential role for mound formation. In Fig.~\ref{fig6a}, 
we investigate the relation between step barriers for bonded and
free particle by analysing the interface width against time at a growth
temperature of $T=573$~K and a fixed step barrier for free particles
$E_{b0}=0.06$~eV. We define a relative step barrier as $\Delta E_b =
E_{b0}-E_{bn}$. When the relative barrier is positive, mounded morphologies with
a large interface width are obtained as can be seen in the bottom snapshots of
Fig.~\ref{fig6b}. For a negative relative barrier, the surface has a long phase,
lasting for several monolayers, with an almost layer-by-layer growth but the
kinetic roughening is still observed for sufficiently long times. Videos 3 and 4
of the supplementary material show the surface evolution for negative and positive
relative barriers. The kinetic roughening regime is absent in the extreme case
of a forbidden interlayer diffusion of the bonded particles ($E_{bn}=\infty$) 
as shown in the most top left snapshot of Fig.~\ref{fig6b} and in Fig.~\ref{fig6a}. 
Notice that varying the relative step barrier  from -0.02 to
0.04~eV, we were able to reproduce a rich variety of spatio-temporal behaviours
ranging from layer-by-layer to self-assembled three-dimensional structures.  Finally, 
it is worth noticing that the morphology dynamics does not change considerably
for low temperatures when the relative step barrier is varied (data not shown)
since, in theses cases, the diffusion of bonded particles is itself negligible.

\begin{figure}[t]
 \centering
 \includegraphics[width=9.5cm]{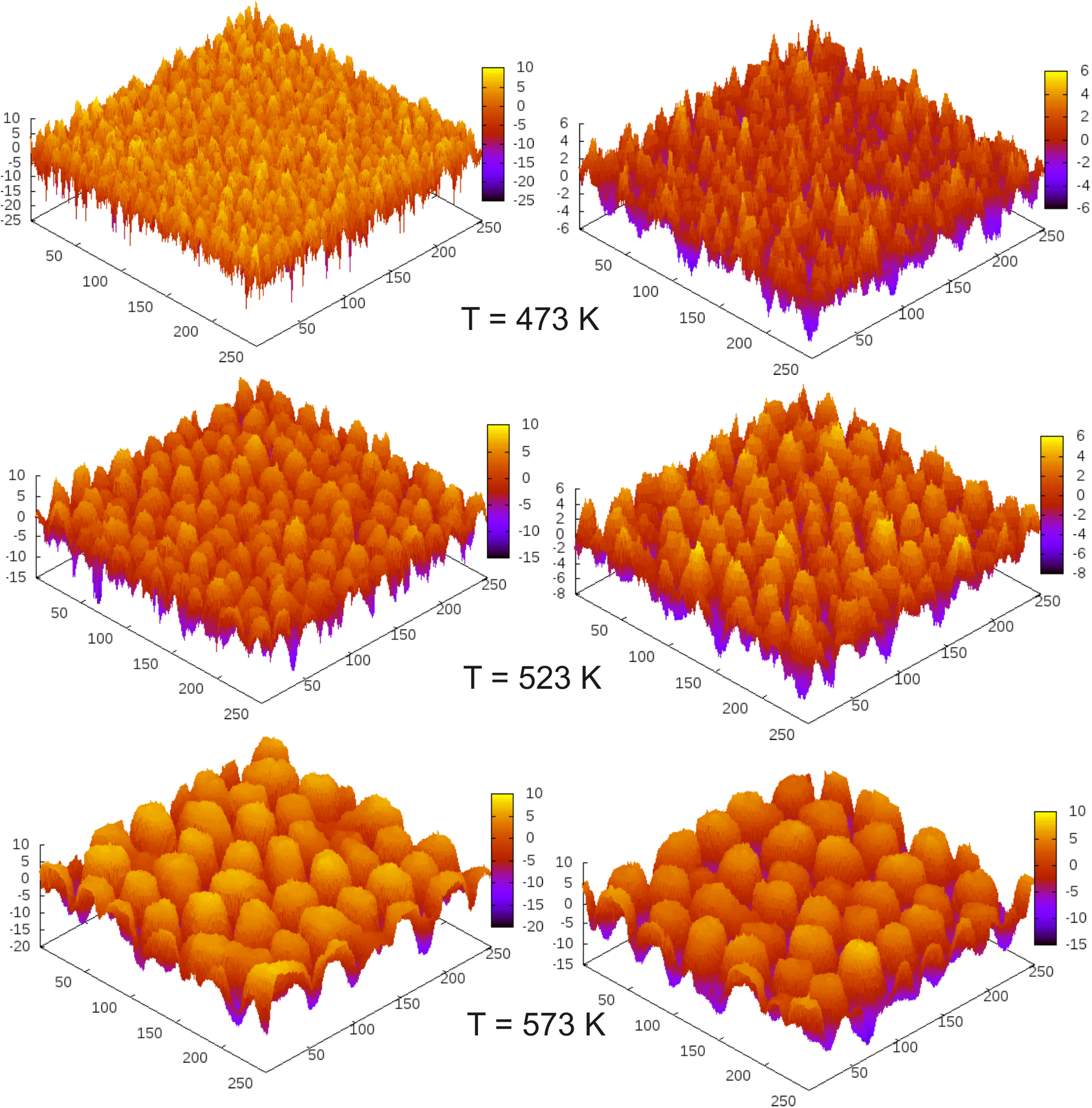}
 % WVvsSOS.pdf: 546x545 pixel, 72dpi, 19.26x19.23 cm, bb=0 0 546 545
 \caption{Surfaces at different temperatures obtained after a deposition of 100~ML
for the simple SOS (left panels) and WV (right panels) deposition rule.}
 \label{fig7}
\end{figure}

In principle, our model has two competing mechanisms for mound
formation: the deposition with a transient mobility and the thermally
activated diffusion. However, the transient mobility contribution to mound
formation is very weak as shown for the WV model ~\cite{Chatraphorn} and becomes
negligible when compared with the diffusion contribution at temperatures
investigated in the present work. However, the transient mobility mechanism is still
important to prevent anomalous grooves in surface, particularly, at the lower
investigated temperatures. In Fig.~\ref{fig7}, we compare surface morphologies
generated using a transient mobility deposition rule with simulations using a
simple SOS deposition, where the particles are deposited directly in a
randomly selected site. Notice that both models generate mounded surfaces even
for a small amount of deposited material. As one can see, the surfaces are
equivalent at higher temperatures, but at lower temperatures, the surfaces for a
simple SOS deposition are featured by grooves that do not correspond to real
epitaxial systems. In summary, mound formation is controlled by the step
diffusion rules, but the transient mobility in the deposition works as an
important smoothing mechanism at lower temperatures.

\section{Conclusions}
\label{sec:conclu}

We have investigated the mound formation in kinetic Monte Carlo simulations of a
model for epitaxial film growth with thermally activated diffusion. The model
has a step barrier, generically known as Ehrlich-Schwoebel
barrier~\cite{ehrlich:1966,schwoebel:1966}, for interlayer diffusion depending
on the number of bonds of the adsorbed particle. The step barriers are present
for interlayer diffusion when a particle is moving up- or downwardly. Moreover,
the rules does not explicitly distinguish between up- and downward diffusion.

Simulations of our bond dependent step barrier (BDSB) model were compared with the
standard model having only a downward step barrier (DSB). We have shown that DSB
and BDSB are equivalent at low temperatures, when the interlayer diffusion of
bonded particles is small. However, if we increase the substrate temperature in
a range compatible with several experimental situations (typically $\Delta
T\gtrsim100$~K), the surfaces obtained  with the standard model become
unrealistic, forming columnar structures with anomalously huge interface widths
and step heights. Our model, on other hand, generates well-behaved mounded
morphologies, considering a null step barrier for bonded particles. The analysis
of the interlayer current shows that DSB has an anomalously large positive
(upward) current responsible by the unstable columnar growth, whereas the
current in the BDSB model tends to a steady state of balance theoretically
required for mound slope selection \cite{SiegertPRL1994,SchinzerEJB2000}. When
we consider our model with barriers only in descending steps,
we observed regular mounded morphology whenever the barrier for bonded particles
is smaller than that for free particles. This important result suggests that
different barriers for bonded and free particles may be present in real
systems.

The case of equal barriers for bonded and free particles in BDSB model was also
investigated. This situation is particularly interesting due to the symmetry
between down- and upward step barriers (Fig. \ref{fig:model1}). In odds with
previous reports for Wolf-Villain model  with an ES barrier in $d=1+1$
dimensions~\cite{Rangdee2006}, the BDSB model also produces mounds for a
symmetric step barrier. A negative interlayer current algebraically approaching
a null value (power law decay) was observed, meaning that the balance between
down- and uphill diffusion  will be
reached only at infinite times when the interface widths are already saturated.
We have determined the evolution of the interface width and characteristic length in
order to characterize the kinetic roughening regimes. The growth and coarsening
exponents observed were $\beta\approx 0.31$ and $\zeta\approx 0.22$,
respectively. These exponents correspond to the so-called super-roughness
scaling regime where $\alpha=\beta/\zeta>1$.

In the asymmetric case, distinct relative barriers $\Delta E_b = E_{b0}-E_{bn}$
were used. For a negative relative barrier, a long phase with a layer-by-layer
growth is observed at early times and a crossover to a kinetic roughening
emerges at long times. Nearby this crossover, a re-entrant dependence of the
interface width with temperature is observed. This behaviour was also
experimentally observed in the growth of Ag/Ag(100)~\cite{Stoldt2000}, among
other systems~\cite{Evans2006}.

\section*{Acknowledgments}
This work was partially supported by the Brazilian agencies CNPq and
FAPEMIG.  SCF thanks the kind hospitality at the Departament de
F\'{\i}sica i Enginyeria Nuclear/UPC.

\section*{References}
%\bibliography{mounds}
\providecommand{\newblock}{}

\end{document}